# A Simple Incremental Modelling of Granular-Media Mechanics

# P. Evesque

Lab MSSMat, UMR 8579 CNRS, Ecole Centrale Paris
92295 CHATENAY-MALABRY, France, e-mail evesque@mssmat.ecp.fr

### Abstract:

*It is proposed a simple non linear incremental modelling of the compression of granular materials and powders. It describes the main features of undrained compression and of oedometric compression using the results of compression tests at constant $\sigma_2 = \sigma_3$ . It generalises the concept of characteristic state. It shows that trajectories arriving at a given critical state shall pertain to a single surface in the (v,q,p) space, which confirms the recent result demonstrating that Hvorslev's and Roscoe's surfaces are part of the same surface. Effects of induced anisotropy are discussed; comparison with classical camclay model is performed.*

_________________________________________________________________________

In previous papers [1, 2], the author has proposed a simple incremental modelling to describe the mechanics of granular materials. It concerns only i) materials built isotropic and ii) simple loading paths; for instance cyclic effect cannot be described quantitatively within this simple scheme, even if this modelling can be used to get the qualitative behaviours.

However, this modelling has been developed progressively as a function of the research progress; so, the presentation it was given in these papers [1,2] could suffer of stressing too strongly some technical and/or theoretical points and disregard some others. The aim of this paper is to give a global overview of the approach and to discuss the assumptions at their right level. For instance, after few oral presentations it seems that the Rowe's law [3] which is assumed in [1,2] and which relates the stresses to dilatancy has been understood as quite necessary and as one of the main theoretical basis of papers [1,2]. This is not true, since one needs only that it is observed experimentally just approximately during a triaxial compression performed at $\sigma_2 = c^{ste}$. Papers [1,2] require only that the pseudo Rowe's relation which has to be used does not imply the existence of a plastic flow rule. So, in this new paper, it is proposed a synthesis which discusses more deeply the assumptions made in [1,2]. The modelling is presented in a different way which gives a better insight to the hypotheses made. The starting point of the approach is the experimental evidence of existing correlations between stress-strain and volume-strain curves (i.e. the Rowe's law); it is interpreted as implying the existence of a dissipation function which depends ***on the stress field and on dilatancy*** [4] . As most of the calculations have been developed already in previous papers [1,2,4], their results will be just recalled here; we hope this will simplify the understanding. Comparison to the Camclay model is given in the appendix.

In [4], the author and Stéfani have assumed that the dissipation $\delta W$ per unit volume of a sample of granular material could be written as $\delta W = D_{plast} \delta \varepsilon_1$ , with $D_{plast}$ be a function $f(\sigma_1, \sigma_2, K)$ of the stresses $\sigma_1$, $\sigma_2 = \sigma_3$ and of the dilatancy $K = -\partial \varepsilon_v / \partial \varepsilon_1$ (where $\varepsilon_v=\varepsilon_1+\varepsilon_2+\varepsilon_3$ and $\varepsilon_1$ are the volume and axial deformation). In this case these authors have demonstrated that one should observe some correlations between the $(\sigma_1-\sigma_2)/\sigma_2$ vs. $\varepsilon_1$ and $\varepsilon_v$ vs. $\varepsilon_1$ curves during a simple compression triaxial test whose sample remains homogeneous. These main correlated features are for a given granular material [4]:

    i) the system is contracting at the origin of test (isotropic stress field),

    ii) since $\delta W=(\sigma_1\varepsilon_1+\sigma_2\varepsilon_2+\sigma_3\varepsilon_3)$, $D_{plast}$ can be determined from experimental curves

    iii) unique ratio $M=(\sigma_1-\sigma_2)/\sigma_2$ when K=0, (i.e. existence of the characteristic state)

    iv) unique $(\sigma_1-\sigma_2)/\sigma_2$ ratio at large deformation (i.e. this defines the critical state value $M=(\sigma_1-\sigma_2)/\sigma_2$),

    v) unicity of dilatancy at the origin under isotropic stress condition, i.e. whatever $\sigma_2$ and density

    vi) the maximum of stress ratio corresponds to the maximum of dilatancy K .

These features are observed experimentally indeed. Furthermore, [4] gave few possible examples and mentioned that Rowe's relation [3] (i.e. $\sigma_1/\sigma_2 = (1+K)\tan^2(\pi/4+\varphi/2) = (1+M)$ $(1+K)$) is obtained if $D_{plast,Rowe}= M\sigma_2(1+K)$; Schofield & Wroth model [5] is obtained when $D_{plast,Schofield}= (M''/3)(\sigma_1+ \sigma_2+ \sigma_3)(1+K/3)$. In order to get the same critical state (K=0), M'' and M shall be related: M''=3M/(M+3)). So, this theoretical hypothesis $D_{plast}= f(\sigma_1,\sigma_2,K)$ is compatible with most experimental results and agrees with most theoretical approaches. An attempt of a similar description can be found in [6], where few of these six results were got also.

Anyway, using point (ii), one can deduce which relation $f(\sigma_1,\sigma_2,K)$ is really compatible with experimental data. So, experimental data [3,7] show that Rowe's relation is approximately valid so that they lead to consider $f(\sigma_1,\sigma_2,K) = M\sigma_2(1+K)$ as a good approximation.

But what does this equation mean exactly? This is a difficult point: Consider first the way Schofield and Wroth has written their equations (see Eqs 5.4, 5.5, 5.8, 5.9, 5.12 & 5.13 of ref [5]); they define the incremental strain by $\delta\varepsilon=2(\delta\varepsilon_1-\delta\varepsilon_2)/3$ so that the external work is $dW=\sigma_1\delta\varepsilon_1+\sigma_2\delta\varepsilon_2+\sigma_3\delta\varepsilon_3=(\sigma_1-\sigma_2)\delta\varepsilon +(\sigma_1+\sigma_2+\sigma_3)\delta\varepsilon_v/3$ and write the energy of plastic deformation as $(1/v)dW/d\varepsilon=M'p$ . In this case, it turns out that the energy of plastic deformation per unit deformation M'p is independent of the way the system deforms; so there are two possibilities: i) this dissipation can be due to a single perfect plastic mechanism, as these authors have assumed; in this case the flow rule is given by K, which becomes a function of q and p only. ii) However, this energy M'p can be due to a complex plastic process, i.e. plastic mechanism with multiple mechanisms, so that K depends also on the incremental path ds; for instance, let us write the dissipated energy of Schofield & Wroth model as $D_{plast,Schofield}= M''( \sigma_1+ \sigma_2+ \sigma_3)(1+K/3)$ as it is assumed above, this implies that it depends on K; in this case, K shall be considered as a free parameter which varies independently of $\sigma_1$ and $\sigma_2$; then, K cannot be understood as a flow rule corresponding to a plastic mechanism; but it shall be understood in the frame work of the incremental modelling, for which $\delta\varepsilon_v$ and $\delta\varepsilon_1$ shall both depend on the stress increments $\delta\sigma_1$ and $\delta\sigma_2$, so that $\delta\varepsilon_v/\delta\varepsilon_1$ depends on $\delta\sigma_2/\delta\sigma_1$, which is a free parameter compared to $\sigma_1$ and $\sigma_2$.

So, we are faced to choose one or the other of these two possibilities.

**We use the incremental modelling now on** and will discuss in the appendix of the paper its differences with the behaviours predicted by the Schofield & Wroth approach [5]. We assume the strain response $\underline{\delta\varepsilon}$ to a stress increment $\underline{\delta\sigma}$ to be linear by zone on $\underline{\delta\sigma}$ for sake of simplicity [8, 9]. In this case, considering an axi-symmetrical triaxial test which keeps constant the directions of principal stresses, the most general manner of writing the incremental stress-strain relation inside a linear zone is:

$$\begin{pmatrix} d\varepsilon_1 \\ d\varepsilon_2 \\ d\varepsilon_3 \end{pmatrix} = -C_o \begin{pmatrix} 1 & -n' & -n' \\ -n & a & -n'' \\ -n & -n'' & a \end{pmatrix} \begin{pmatrix} ds_1 \\ ds_2 \\ ds_3 \end{pmatrix} \qquad (1)$$

The constants $C_o$, $\alpha$, $\nu,\nu',\nu''$ depend *a priori* on the applied stress $(\sigma_1,\sigma_2=\sigma_3)$, on the history and on the zone. In the case when the granular material is isotropic, its response shall be isotropic, so that Eq. (1) reduces to:

$$\begin{pmatrix} d\varepsilon_1 \\ d\varepsilon_2 \\ d\varepsilon_3 \end{pmatrix} = -C_o \begin{pmatrix} 1 & -n & -n \\ -n & 1 & -n \\ -n & -n & 1 \end{pmatrix} \begin{pmatrix} ds_1 \\ ds_2 \\ ds_3 \end{pmatrix} \qquad (2)$$

Eq. (2) shall apply at the beginning of any triaxial test since at this stage, the samples are assumed to be homogeneous and their contact distribution isotropic; Eq. (2) shall remain true as far as the contact distribution does not evolve significantly; so, it shall remain valid as far as the sample deformation remains small, since the contact distribution can change only if grains move compared to one another.

We assume now on that the domain of a incremental linearity at a given stress field $(\sigma_1,\sigma_2=\sigma_3)$ is large enough to contain all the different compression tests (i.e. compression at $\sigma_2=\sigma_3=c^{ste}$, at $\sigma_1+\sigma_2+\sigma_3=c^{ste}$, undrained compression ($v=c^{ste}$), oedometric compression ($\varepsilon_2=\varepsilon_3=c^{ste}$)). This means *a contrario* that extension tests pertain to one or few other zones, governed by linear relationships similar to Eq. (1) but with different coefficients; this is a simple way to reproduce hysteresis behaviours and history-dependent responses which characterise the behaviour of granular media.

A triaxial cell compression at constant stress $\sigma_2=\sigma_3$ allows to determine the variations of $\varepsilon_1$ and $\varepsilon_v$ with $\sigma_1$. Then it allows to determine the variations of $C_o$ and $\nu$ with $\sigma_1/\sigma_2$. For instance Eq. (1) leads to $\delta\varepsilon_v/\delta\varepsilon_1 =1-2\nu=K_{\sigma_2=c^{ste}}$. In the same time, modelling shall satisfy the Rowe's equation as found from experiment at $\sigma_2=\sigma_3=c^{ste}$. So we get from experiment $\sigma_1/\sigma_2 = (1+M)(1+K_{\sigma_2=cste})$ ; this leads to $\sigma_1/\sigma_2 = (1+M)(2-2\nu)$. This can be rewritten:

$$\nu = 1 - \sigma_1/\{2(1+M)\sigma_2\} \qquad (3)$$

One can remark that Eq. (3) imposes $\nu=1/2$ when $q=(\sigma_1-\sigma_2)=M\sigma_2$. We will call this state the characteristic state because it does not change of volume whatever the increment of stress [10]. It is characterised by $K_{\sigma_2=c^{ste}}=0$, or $\nu=1/2$) and by $q=M\sigma_2$. This relation is valid for any dense or loose material, and whatever the isotropic or anisotropic nature of the response.

In principle, to determine the values of the other parameters $\alpha$, $\nu'$, $\nu''$ , one needs to use other paths pertaining to the same incremental zone. This requires that the incremental zone is large enough. So, in the next it will be assumed that the incremental domain is large enough to contain triaxial compression test at constant volume (undrained test or $v=c^{ste}$), at constant radius ($\epsilon_2=\epsilon_3=c^{ste}$), at constant mean pressure $(p=[\sigma_1+\sigma_2+\sigma_3]/3=c^{ste})$ and at constant lateral stress $(\sigma_2=\sigma_3=c^{ste})$ . Furthermore, as the number of existing different tests is limited one shall expect that the modelling works without needing too many adjustable parameters. To reduce the number of parameters, we will always start and consider an isotropic modelling; we will discuss what this modelling describes and what it does not, and how the misfit can be understood in term of anisotropy. Let us consider first the case of an undrained test.

## Undrained ($v=cste$) behaviour:

In the case of undrained results, $\epsilon_v=0$; so typical results can be summed up knowing the trajectory in the $(q=\sigma_1-\sigma_2=, p=[\sigma_1+\sigma_2+\sigma_3]/3)$ and knowing the amplitude of the deformation $\epsilon_1=$ as a function of $q=\sigma_1-\sigma_2$. Typical results are reported in Fig. 1.

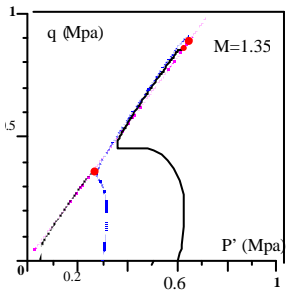

**Figure 1:** Typical experimental results of undrained test on Hostun sand (after Flavigny).

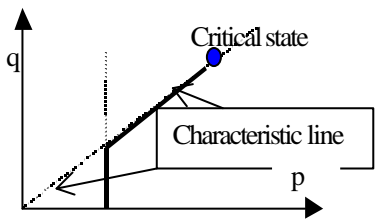

**Figure 2:** calculated trajectories of undrained test on a dense sample. In the q,p plane, the path starts vertical (p=cste). A trans-critical bifurcation occurs when the trajectory meets the $q=M\sigma_2$ line. And the path follows this line till it meet the critical state line.

In [2], the behaviour has been calculated assuming that the medium remains isotropic. This is valid till deformations remain small. Hence, it shall be true for dense materials far enough from the critical state. So, for dense enough materials one shall expect that Eq. (2) models the behaviour during the vertical path and part of the $q=M\sigma_2$ line. In turn, this imposes that the isotropic modelling used in ref. [2] is valid for dense materials. In [2], it has been demonstrated that the angle observed when the trajectory arrives at the line $q=M\sigma_2$ corresponds to a transcritical bifurcation which is imposed by the condition $\delta\varepsilon_v=0$ which becomes satisfied whatever $\delta\sigma_1$ and $\delta\sigma_2$ when $q=M\sigma_2$ since $\nu=1/2$ (see Eq. (3) above), and because the solution $\delta\sigma_1-\delta\sigma_2 =M\delta\sigma_2$ which is now possible dissipates an energy smaller than the one which imposes $\delta p=0$. Conversely, one can consider the argumentation of [2] correct and that $\nu=1/2$ when the sample remains isotropic and on the characteristic line, i.e. when $q=M\sigma_2$ that is to say.

***Discussion:*** However, one observes that the trajectory follows the same $q=M\sigma_2$ characteristic line when the sand is initially less dense, and/or when it is largely deformed. But in this case the response cannot be considered as isotropic anymore and one shall use the general modelling of Eq. (1). This implies that:

$$(1-2\nu)\delta\sigma_1+2(\alpha-\nu'-\nu'')\delta\sigma_2=0 \qquad (4)$$

Since $q=M\sigma_2$ on the characteristic line, $\delta\sigma_1=(M+1)\delta\sigma_2$; hence, Eq.(4) leads to the general relationship for the post-bifurcated undrained path:

$$(1-2\nu)(M+1)+2(\alpha-\nu'-\nu'')=0 \quad \text{when } q=M\sigma_2 \qquad (5)$$

Furthermore, if one assumes that Eq. (3) holds true also since $q=M\sigma_2$, this imposes $\nu=1/2$; and one gets from Eqs. (4 & 5):

$$\alpha-\nu'-\nu''=0 \quad \text{when } q=M\sigma_2 \qquad (6)$$

But this demonstration requires that Eq. (3) holds true here. This is plausible since it holds i) for the classical triaxial test whatever the sample anisotropy and ii) for the undrained test as far as the sample remains isotropic ($\alpha=1$, $\nu'=\nu''=\nu$). Under this hypothesis, Eq.(3) and Eq. (6) seem to characterise the mechanical response of the characteristic state. But further work is required to confirm this model. Anyway, either Eq.(5) or the combination of $\nu=1/2$ and Eq. (6) corresponds to a new definition of the characteristic state.

At last, in Fig. 1, one observes that the trajectory is not perfectly vertical at the beginning; this can be due to two different causes: either it may be due to not perfectly undrained conditions, caused by a compressible fluid which fills the pores partly; in this case one observe in general a slope $q=3p$ at the beginning as in a triaxial test at constant $\sigma_2=\sigma_3$. Or it can be induced by existing residual anisotropy which has been generated during the building process; if so, this anisotropy is small however, since the path is only slightly inclined.

Finally, the fact that the trajectory starts and remains vertical in the $(q,p)$ plane till it reaches the characteristic line for dense enough piles demonstrates that the **mechanical response remains isotropic till deformation has not proceeded importantly**. In other words, this implies that the **system does not develop stress induced anisotropy** alone (i.e. without deformation). So, in this granular-matter systems, anisotropy is induced by deformation which allows a change of the contact distribution; it seems that it cannot be induced by a change of the stress distribution which does not generate deformation. Furthermore, writing $d\sigma_1=dp+2dq/3$ and $d\sigma_2=d\sigma_3=dp-dq/3$, Eq. (1) leads to the undrained condition:

$$(1+\alpha-2\nu-\nu'-\nu''){3dp+2(1-\alpha-2\nu+\nu'+\nu'')dq}=0 \qquad (7)$$

which reduces to dp=0 or to 1-2ν=0 in the isotropic case (i.e. $\alpha=1,\nu=\nu'=\nu''$) and which defines the slope dq/dp of the trajectory before it reaches the characteristic line in an anisotropic case, i.e. when $1-\alpha-2\nu+\nu'+\nu''\neq0$ . It defines the characteristic state solution: $1+\alpha-2\nu-\nu'-\nu''=0$ when this solution is possible.

## Oedometric path

In the general case, the condition $d\varepsilon_2=d\varepsilon_3=0$ combined with Eq. (1) imposes:

$$-\nu d\sigma_1+d\sigma_2(\alpha-\nu'')=0 \qquad (8)$$

which imposes the stress increment $d\sigma_2= [-\nu/(\alpha-\nu'')]d\sigma_1$ for a given $d\sigma_1$. Let consider first a sand sample and a dense granular medium; we known that axial deformation is small in such a case so that the contact distribution evolves very little during the test because there is little reorganisation of the grains. So one can consider that the mechanical response shall be isotropic, i.e. $\alpha=1,\nu=\nu'=\nu''$. However, this does not mean that ν has not evolved: Turning back to the case of a triaxial test at $\sigma_2=\sigma_3=c^{ste}$ on a dense sand sample, very little deformation is observed also before the sample reaches its minimum of volume; in this case also the contact distribution has little evolved; but its pseudo Poisson coefficient has evolved and it obeys a stress dependence (i.e. the Rowe's law).

So, combining Eq. (8) with $\alpha=1,\nu=\nu'=\nu''=$ and with Eq. (3) (i.e. $\nu=1-\sigma_1/{2(1+M)\sigma_2}$) leads to the stress evolution equation:

$$d\sigma_2/ d\sigma_1=\nu/(1-\nu)=[2(1+M) \sigma_2-\sigma_1]/\sigma_1 \qquad (9)$$

This equation has been integrated in [1]; it has been demonstrated that it converges fast towards an asymptotic ratio of $(\sigma_2/\sigma_1)_{oed} \cong 1-\sin\varphi$ which depends on M only, so that this asymptotic ratio is approximately the Jaky constant 1-sinφ.

## *At this stage few remarks are worth to be done:*

1) The exact asymptotic ratio $(\sigma_2/\sigma_1)_{oed}$ depends on the exact dependence of ν upon $(\sigma_2/\sigma_1)$. It might occur that the Rowe's law is not followed exactly by the sample; in this case, the exact asymptotic value $(\sigma_2/\sigma_1)_{oed}$ will change ; however, such a ratio $(\sigma_2/\sigma_1)_{oed}$ will exist as far as it exists some relationship $\nu=g(\sigma_2/\sigma_1)$ between ν and $(\sigma_2/\sigma_1)$; this asymptotic value is obtained by integrating Eq. (9).

2) This is probably the case of Hostun sand, for which it has been observed that the experimental dilatancy during a $\sigma_2=c^{ste}$ test deviates from that one of the Rowe's law in the range between $0.6<\sigma_2/\sigma_1<1$ and is only satisfied below 0.6; however, the value of K (and ν) depends also on $\sigma_2/\sigma_1$ in this range (0.6,1) so that the $\nu=g(\sigma_2/\sigma_1)$ law has just to be modified. Integration of Eq. (9) is still possible and one shall find a defined asymptotic value. The asymptotic value might occur to be quite different if ν varies too steeply during $1>\sigma_2/\sigma_1>0.6$; however, in the present case of Hostun sand, it should remain identical to the predicted one since Rowe's relation is satisfied in the region interesting the asymptotic behaviour ($\sigma_2/\sigma_1 < 0.6$).

3)   As the modelling leads to an accurate description of the $(\sigma_2/\sigma_1)_{oed}$ ratio, the isotropic response assumption seems to be valid. However, let us assume that it is not valid; this is still possible if $(\alpha-v'')=1-v$  so that the general Eq. (8) (i.e. -$vd\sigma_1+d\sigma_2(\alpha-v'')=0$) leads to the same result for the oedometric test on sand.

4)   Furthermore, let us turn back to the undrained test; in this case, Eq. (1) applied to undrained condition leads to  $d\epsilon_v=(1-2v)d\sigma_1+2d\sigma_2(\alpha-v'-v'')=0$ . So, combining remark (3) (i.e. $\alpha-v''=1-v$) with the experimental observation that dp=0 during the first stage of the undrained test, one can conclude that $v=v'$ during this step.

5)   It is worth noting that in an axi-symmetric triaxial test the number of possible experiments is not enough to allow the determination of $\alpha$ and $v''$ separately, so that one can choose arbitrarily $\alpha=1$, without loss of generality. In this case remarks (3,4) implies that  $v=v'=v''$. This means that for any  axi-symmetric triaxial test on isotropic packing whose packing distribution of contacts remain isotropic, the mechanical response can be considered as isotropic with a pseudo Poisson coefficient which depends on  $\sigma_1/\sigma_2$, as far as the  undrained test path occurs at dp=0 at least; this hypothesis might still be true on the first part of the q=Mp line if the path has not deviated from dp=0 before reaching this q=Mp line.

6)   This model is valid for oedometer test on sand; however it is worth mentioning that its application to clay appears to be valid experimentally; this might be more difficult to understand because deformations are much larger there, so that induced anisotropy of contacts should be larger, leading to a larger  anisotropic domain.

7)   Application to natural soils: It is worth recalling that the modelling presented here is only valid starting from an isotropic pile and increasing the  deviatoric stress. Due to that, it cannot apply directly in many circumstances in nature: for instance, the stress relation of natural sand layers do not obey the oedometric behaviour very often; this is caused by the anisotropy of the sand layer during sedimentation due to water stream,…. In the case of clay this stress relation is more often observed; however, the stress shall have increased continuously all over the years, otherwise the system would not be in constant compression and could be in extension, and the model does not apply due to memory effects.

8)   In the case of laboratory oedometer tests, the described behaviour can be observed during the first compression; it cannot be observed during axial unloading.

## Hvorslev-Roscoe

In [11], it has been proposed to understand the topology of the trajectory arriving at the critical state in the phase space (q,p,v) using the qualitative theory of dynamical systems. This has led to the conclusion that Hvorslev's and  Roscoe's surfaces shall be part of the same 2d regular surface. Furthermore, we have shown that experimental data confirm this modelling: trajectories arriving from dense and loose packing arrive just in opposite directions at the critical state, and this for two different and independent sets of tests (undrained tests and $\sigma_2=\sigma_3=c^{ste}$ compression).

We want to confirm this result using an other approach, the incremental modelling. As a matter of fact, this modelling obeys Eq. (1). In Eq. (1) the number of free independent parameters is 2, i.e. $d\sigma_1$, $d\sigma_2=d\sigma_3$. Now, considering the matrix representing the mechanical behaviour at the critical state, it shall be uniquely defined, since the mechanics of the critical state is well defined; so it defines a 2d surface, which is just the surface of the trajectories arriving at the critical state; this surface shall then contain the Hvorslev's and Roscoe's surfaces, whose tangent at the critical state pertain to the same plane. This is just what we wanted to demonstrate. This analysis does not depend on the fact that the response is isotropic or not.

However, in the case when real 3d experiment would be performed ($\delta\sigma_1\neq\delta\sigma_2\neq\delta\sigma_3$), they would impose real 3d trajectories since the number of independent increments would be 3. So, we see that the notion of Hvorslev's and Roscoe' surface is limited to axisymmetric problems, and is perhaps meaningless.

**Discussion:**

*1) Is the matrix M governing Eq. (1) symmetric?*

In [1,2] this was assumed since $v=v'$ there; in this paper, such an hypothesis is not assumed. It can be shown that this assumption is needed if the system is reversible; it is known that mechanics of powders and granular matter is not reversible, so this hypothesis has to be verified from experiments. However this has not much consequences about the main results of papers [1,2], as it is discussed in the present paper.

*2) effect of a spontaneous breaking of the axial symmetry*

In many experimental test, one observes the formation of a spontaneous yield surface which breaks spontaneously the axial symmetry of the mechanical response so that one should introduce two coefficients for $v$, $v'$, $v''$ and $\alpha$, one for each direction $x_2$ and $x_3$. However, as one imposes always the same values of $\sigma_3=\sigma_2$ and of $d\sigma_3=d\sigma_2$, one shall expects that the above analysis remains true as far as the stress field remains approximately homogeneous in the sample. It requires just to used average quantities.

*3) the isotropic modelling: a simple efficient modelling*

It is worth mentioning that the isotropic modelling of Eq. (2) is able to describe most of the main characteristics of the mechanics. Despite this, it is not able to describe the decrease of mean pressure during an undrained test for looser materials, which is generated by the development of strain-induced anisotropy.

It is remarkable that characteristic state behaviour does not depend on the amplitude of strain-induced anisotropy, so that the undrained path follows always the same characteristic line when the trajectory has reached it, whatever the induced anisotropy. (It is worth noting however that in some circumstances the trajectory can overpass slightly the $q=M\sigma_2$ line when it reaches it, but it comes back on it very quickly).

### *3) mean field approach: a difficult approach*

It is worth recalling the difficulty one can meet when using mean field approach; this can be illustrated by the Rowe's law and its story: Rowe has deduced his law by considering regular ordered arrays of cylinders; he has remarked that grains can slide only in particular directions which depend on the lattice orientation and he has obtained a plastic flow rule; however, he has been able to write this flow condition of sliding in such a way that it looks like not to depend anymore on the lattice structure, but only on the stress field. As this law is valid for any sliding lattice, it seems to be valid after averaging over any set of configuration and Rowe has concluded that his relation was quite general. ); this relation has been called the Rowe's relation.

Nevertheless, this averaging is not possible since the relation is only valid for those lattices which are at sliding and not for the others. In other words, Rowe method assumes implicitly that his relation applies to both activated and not-activated sliding mechanisms at the same time, which is not true. This shows the difficulty of interpreting a mean field approach and to prove its exactness.

### Conclusion:

One of the main conclusion of this paper is that an isotropic description allows to reproduce most of the mechanical behaviours under simple compression paths: it allows to describe in a simple way typical oedometric and undrained compressions from typical axisymmetric triaxial compression at $\sigma_2=\sigma_3=c^{ste}$. It can describe also the way the material arrives at the critical state, that is to say, it contains the existence of the Hvorslev's and Roscoe's surfaces.

The second conclusion is that the anisotropy of the mechanical response develops only during deformation. In counter part, stress loading without generating deformation does not generate an anisotropy of the mechanical response.

However, anisotropy can develop too fast in the case of loose or slightly dense materials, so that experimental behaviour deviates from the present modelling noticeably. In this case the adequate modelling shall take into account anisotropy.

An other source of error is the development of heterogeneity. This can arrive spontaneously as in the case of the localisation of the deformation (shear banding), or in the case of spontaneous barrelling bifurcation [12].

### APPENDIX: Comparison with camclay model

The cam clay model is an elasto-plastic model with a single plastic mechanism. The yield curve depends on the density of the material and the flow rule is given by [5]:

$$pdv/v +qd\varepsilon=M''p|d\varepsilon| \text{ , with } d\varepsilon=2(d\varepsilon_1-d\varepsilon_2)/3 \qquad (10)$$

so that $(dv/v)/d\varepsilon$ is a unique function of the stress field (q,p). Plasticity theory applied to Eq. (10) leads to a series of yield curves, one of which is sketched on Fig. 3. Using this modelling, one can find predicted behaviours and compare them to experiments:

## *1)  Triaxial compression at $s_2=s_3=$cste:*

The path corresponds to a line inclined at $dq=3dp$ in the $(q,p)$ plane. Starting from $q=0$, the material behaves first elastically with a Young modulus and a Poisson coefficient. Hence, its trajectory in the $(q,p)$ plane is $p=q/3+p_o$. Then the trajectory reaches the yield curve; at this stage it becomes plastic so that it shall expand or contract during deformation; the volume change is given by Eq. (10) and depends on $(q,p)$. Anyhow, its trajectory obeys $p=q/3+p_o$ which is the experimental test condition.

If the material is dense initially, its flow rule exhibit dilatancy when reaching the yield curve (i.e. $p_o<(1+M''/3)p_u$); so its specific volume shall decrease and the yield curve shrinks; the trajectory comes back on the $p=q/3+p_o$ line; this processes continues till the trajectory reaches the $q=M''p$ line for which there is no more volume change whatever the deformation.

On the contrary, in the case of a loose sample (i.e. $p_o>(1+M''/3)p_u$), the sample contracts during plastic deformation so that the yield curve expands; the trajectory follows the $p=q/3+p_o$ line towards increasing p till it reaches the $q=M''p$; at this stage the material deforms plastically without volume variation and without evolution of the yield curve. This is the critical state.

So the model predicts very little deformation at the beginning (i.e. small $q/p$ ratio) and elastic behaviour (i.e. reversibility); both facts are not observed experimentally. It does not respect the Rowe's relation. However, it captures the dilatant or contractant characteristics at large stress depending on the $q/p$ ratio.

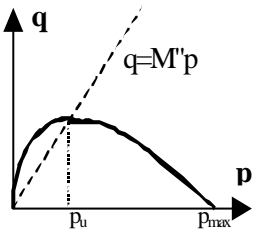

**Figure 3:** The yield curve of the Cam clay model for a given specific volume $v_1$. It obeys the equation:

$$q/M''p + \ln(p/p_u)=1 \qquad \text{for } q>0 \qquad (11)$$

The maximum value of p of this curve is $\ln(p_{max}/p_u)=1$; the maximum of q occurs at $p_u$. The line $q=M''p$ is the characteristic line (no volume change). Volume change is governed by Eq. (10) on this curve. The material has an elastic behaviour inside the curve and a plastic one when reaching the yield curve. The specific volume $v_l$ corresponding to this yield curve is $v_l=\Gamma-\lambda \ln p_u$, where $\Gamma$ and $\lambda$ are two parameters. The large $p_u$, the smaller $v_1$.

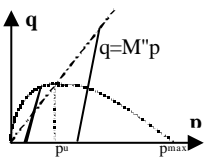

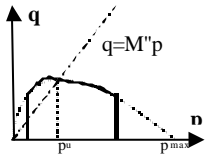

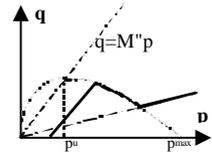

**Figure 4:**

a)   triaxial test at $\sigma_2=\sigma_3=$cste

b) undrained test

c) oedometric test ($\varepsilon_2=\varepsilon_3=0$) in the plastic region the path is governed by $q=(M''-1)p$

## *2) Undrained (* **v=cste** *) compression:*

Let us first consider that q=0 and p>$p_u$ at the beginning of the test; this is the case of a loose material. In this case, due to the elastic behaviour, the trajectory in the (q,p) plane starts vertically; during this part the deformation shall remain very small (<$10^{-3}$). Then the trajectory reaches the yield curve on its right part; at this point the material becomes plastic; however, as it cannot dilate or contract due to experimental condition, it shall follow the yield curve without deforming noticeably; this implies that the trajectory turns left on the yield curve and follows it till the stress reaches the q=M"$p_u$ point. At this stage the trajectory shall stop, since plastic flow rule imposes $\delta\varepsilon_v$=0 so that samples cannot contract or dilate and the yield curve cannot evolve. The trajectory shall follow the yield curve till it reaches the q=M"p value where it shall stops because q=M"p imposes $\delta\varepsilon_v$=0. Both facts are not consistent with experimental results.

Let us now consider that q=0 and p<$p_u$ at the beginning of the test; this is the case of a dense material. The trajectory starts vertically as in the preceding case, till it reaches the yield curve; at this location it shall turn right on the yield curve and follows it till the q=M"$p_u$ point where it shall stop.

So, the predicted trajectory should exhibit a bifurcation with a left (right) turn for loose (dense) samples and should stop as soon as q reaches the q=M"p value. Both predictions are not consistent with experimental results.

## *2) Oedometric (* **e₂=e₃=0** *) compression:*

In the case of the oedometric test, the response obeys first an elastic response; so the elastic Poisson $\nu_{elast}$ coefficient imposes the increment of stress ratio

$$d\sigma_2/d\sigma_1 = \nu_{elast}/(1-\nu_{elast}) \qquad => \qquad dq/dp = 3(1-2\nu_{elast})/(1+\nu_{elast})$$

Hence the first part of the trajectory is an inclined line; typical slope is dq/dp≅1.5 since typical $\nu_{elast}$≅0.2. Then the trajectory reaches the yield curve, on which it shall turn either right or left depending on the initial void ratio corresponding to $p_u$. If we now neglects the elatic part of the deformation now on, the trajectory shall follow the yield curve till the dilatancy corresponds to the experimental condition $\delta\varepsilon_2$=0, i.e. $\delta\varepsilon_v$=$\delta\varepsilon_1$ ; since plastic deformation is blocked by experimental condition otherwise. So, imposing $\delta\varepsilon_2$=0 in Eq. (10) leads to p +q=M"p , so that q/p=M"-1(≅0.2 since typical value of M"=1.2 for φ=30°). When this working ratio is reached the trajectory can follow it: increasing q makes the system to contract and the yield curve to expand; this deformation process is compatible with experimental conditions and can proceed. This is sketched on Fig. (4c). As M and M" are related together, i.e. M"=3M/(M+3), a typical value of M" is 1.2 (for φ=30°), a typical value of $\eta_{oedom-camclay}$=(q/p)$_{oedom-camclay}$ is $\eta_{oedom-camclay}$ =0.2, which corresponds to a stress oedometric ratio $K_{oedom-camclay}$ =$\sigma_3$/$\sigma_1$=(4-M")/(2M"+1)=(3-η)/(2η+3)≅0.82, which is much too large.

In conclusion, i) cam clay model requires to introduce the elastic behaviour if one wants to describe the stress evolution when deviatoric stress q is small. However, doing so it does not consider that plastic deformation can occur already at small q and before few percents of deformation. ii) it is not compatible with Rowe's law, since this one is valid for dense materials before the stress peak, when cam clay suppose an elastic response. iii) Cam clay does not get the right oedometric stress ratio. iv) It leads to predict 2 bifurcations for the trajectory in the (q,p) plane in most cases of simple (oedometric, …) compressions; this is not observed in general, since no bifurcation is observed most of the time , except bifurcations linked to developments of inhomogeneity such as localisation, bubble formations, cavitation,…). In the case of undrained test, a single bifurcation is predicted by camclay, but it does not correspond to the one observed experimentally because it does not predict the part of the path along the characteristic line.

This camclay modelling looks then less efficient that the one proposed in this paper.

***Acknoledgments****:* CNES is thanked for partial funding. Discussions with Prof. J. Biarez, R. Chambon, F. Darve, J. Desrues, E. Flavigny, G. Gudehus and A. Moderessi have been appreciated.

The electronic arXiv.org version of this paper has been settled during a stay at the Kavli Institute of Theoretical Physics of the University of California at Santa Barbara (KITP-UCSB), in june 2005, supported in part by the National Science Fundation under Grant n° PHY99-07949.

*Poudres & Grains* can be found at :
http://www.mssmat.ecp.fr/rubrique.php3?id_rubrique=402